\documentclass[a4paper]{article}
\usepackage{amsmath}
\usepackage{latexsym}
\usepackage{graphicx}
\usepackage{mathptmx}
\usepackage{bm}

\title{On a type of non-classical boundary condition of Lagrangian field}

\author{Zaixing Huang\\[5pt]
  State Key Laboratory of Mechanics and Control of Mechanical Structures\\
  Nanjing University of Aeronautics and Astronautics\\
  Yudao Street 29, Nanjing, 210016, P R China\\ E-mail: huangzx@nuaa.edu.cn}

\begin{document}

\maketitle

\begin{abstract}
In the framework of the Lagrangian field theory, we derive a type of
new non-classical natural boundary condition to be correlated with
the mean curvature of boundary surface. Under the condition of
homogeneity and isotropy, this type of boundary condition can be
simplified into the Tolman's formula in which the size effect of
surface tension is prescribed.
\end{abstract}

\section{Introduction}
\label{intro} Conventionally, boundary conditions of partial
differential equation can be categorized into four types: Dirichlet,
Nuemann, Robin and periodic boundary condition. All of these
boundary conditions are determined in terms of a self-adjoint
extension of the differential operator of field, rather than
concerning the surface effects of boundary. With entering
micro/nano-scale, the surface effect has to be taken into account in
the behaviors of material. This causes non-classical boundary
conditions to appear in the boundary value problems of partial
differential equations. Some typical examples can be found in
capillary wave, surface elasticity and phase transition etc.

The first non-classical boundary condition is the Young-Laplace's
equation \cite{1}. As a traction boundary condition, it was used to
solve the oscillation of spherical droplet \cite{2} and fission of
nucleus \cite{3}. Gurtin and Murdoch extended the Young-Laplace
equation into the generalized Young-Laplace equation so as to
characterize the surface of elastic solid \cite{4}. Further,
Steigmann and Ogden proposed reinforced boundary condition by taking
into account the bending stiffness of the surface film \cite{4s1}.
Zhu, Ru and Chen discussed non-uniqueness of boundary value problems
based on the generalized Young-Laplace equation \cite{4s2}. Javili
and Mosler et al revisited and carefully examined the
surface/interface elasticity theory. They established a consistent
linearized interface elasticity theory \cite{4s3}.

As a traction boundary condition, the generalized Young-Laplace
equation has been applied to investigate physical behaviors of
nano-structured materials. The relevant literatures can be found in
the reviews by Wang et al. \cite{5} and Sun \cite{6}. Recently,
Figotin and Reyes advanced a non-classical boundary condition, whose
feature consists in that the boundary fields may differ from the
boundary limit of the interior fields so as to characterize the
interactions between the boundary and the interior fields \cite{7}.
Huang proposed a Shape-dependent natural boundary condition
\cite{8}. However, there is an error in \cite{8} due to mistakenly
using the divergence theorem on surface. So far, many studies have
shown that the influences of surface effect on physical behaviors of
field can be characterized by the boundary condition. However, how
to introduce the surface effect in the boundary condition is still a
problem awaiting to be further explored. So the aim of this paper is
to propose a type of non-classical boundary condition that can
simultaneously characterize the surface effect and its size effect
in the framework of the Lagrangian field theory.

The paper is outlined as follows. In Section 2, we introduce a
surface Lagrangian to describe the surface effect of field. The
Lagrangian equation and curvature-dependent natural boundary
condition are derived. In Section 3, by simplification to the
curvature-dependent natural boundary condition, the Tolman' formula
is given. Finally, we summarize and comment on the results in this
paper.\\

\textbf{Notation:} The index rules and summation convention are
adopted. Latin indices run from 1 to 3. The Greece letter $\Omega$
stands for a bounded domain of $R^3$, and $\partial\Omega$ is the
boundary surface of $\Omega$. The covariant derivative with respect
to coordinates is represented by the symbol $\partial_k$. The
contravariant derivative operator corresponding to $\partial_k$ is
denoted by $\partial^k=g^{kj}\partial_j$, where $g^{kj}$ is the
metric tensor. The symbol $\partial_A$ ($A=1, 2$) or $\nabla_s$ is
the surface gradient operator defined on $\partial\Omega$. The
derivative with respect to time is denoted by an upper dot, e.g.,
$\dot{a}={\textrm d}a/{\textrm d}t$. Other symbols will be
introduced in the text where they appear for the first time.

\section{Boundary condition of Lagrangian field}
\label{sec:1} Let $\textbf{x}=\{x^k\}$ be a 3-dimensional position
vector in $\Omega$ and $t\in[t_0, t_1]$ be time. A vector field
defined on $[t_0, t_1]\cup\Omega$ is denoted by $\phi_k=\phi_k(t,
\textbf{x})$. The Lagrangian of the field $\phi$ is written as
$L=L(\phi_k,\dot{\phi}_k,
\partial_j\phi_k)$.

Let spatial domain $\Omega$ occupied by $\phi_k$ be bounded and the
surface $\partial\Omega$ of $\Omega$ be a smooth surface. We believe
that physical behaviors of $\phi_k$ in the interior of $\Omega$ are
different from those on the boundary of $\Omega$. An additional
Lagrangian $\Gamma$ is used to characterize the physical behaviors
of $\phi_k$ on the boundary surface $\partial\Omega$. We refer to
$\Gamma$ as the surface Lagrangian, which is supposed to have the
form below
\begin{equation}\label{s1}\Gamma=\Gamma_0(\phi_k,\dot\phi_k,\partial_A\phi_k)+
\nabla_s\cdot\mathbf{S}(\phi_k,\dot\phi_k,\partial_A\phi_k).\end{equation}   
On $\partial\Omega$, the vector field
$\mathbf{S}(\phi_k,\dot\phi_k,\partial_A\phi_k)$ can be decomposed
into $\mathbf{S}(\phi_k,\dot\phi_k,\partial_A\phi_k)=
S^A(\phi_k,\dot\phi_k,\partial_A\phi_k)\mathbf{g}_A+\hat{\Gamma}(\phi_k,\dot\phi_k,\partial_A\phi_k)\mathbf{n}$,
where $\mathbf{g}_A$ is the unit base vector defined on the tangent
plane of $\partial\Omega$ and $\mathbf{n}$ the unit normal vector.
As thus, Eq.(\ref{s1}) is rewritten as
\begin{equation}\label{s2}\Gamma=\Gamma_0+\partial_AS^A-2H\hat{\Gamma},\end{equation}        
where $H$ is the mean curvature of $\partial\Omega$. In general, it
is explicitly independent of time. In the process to derive
Eq.(\ref{s2}), we use the identity $\nabla_s\cdot\mathbf{n}=-2H$
\cite{9,10}. By Eq.(\ref{s2}), the action of field can be
represented as
\begin{eqnarray}\label{s3}A[\phi_k]&=&\int_{t_0}^{t_1}\int_\Omega L(\phi_k,
\dot\phi_k,\partial_j\phi_k)\textrm{d}v\textrm{d}t+
\int_{t_0}^{t_1}\int_{\partial\Omega}\Gamma(\phi_k,\dot\phi_k,\partial_A\phi_k)
\textrm{d}a\textrm{d}t\nonumber\\
&=&\int_{t_0}^{t_1}\int_\Omega L\textrm{d}v\textrm{d}t+
\int_{t_0}^{t_1}\int_{\partial\Omega}(\Gamma_0+\partial_AS^A-2H\hat{\Gamma})\textrm{d}a\textrm{d}t\nonumber\\
&=&\int_{t_0}^{t_1}\int_\Omega L\textrm{d}v\textrm{d}t+
\int_{t_0}^{t_1}\int_{\partial\Omega}(\Gamma_0-2H\hat{\Gamma})\textrm{d}a\textrm{d}t,\end{eqnarray}          
where $\textrm{d}v$ and $\textrm{d}a$ are a volume measure in
$\Omega$ and an area measure on $\partial\Omega$, respectively. Let
$\delta\phi_k(t_0)=\delta\phi_k(t_1)=0$. Taking the variation of
$A[\phi_k]$ leads to
\begin{eqnarray}\label{s4}\delta A&=&\int_{t_0}^{t_1}\int_\Omega\{\frac{\partial L}{\partial
\phi_k}-\frac{\textrm{d}}{\textrm{d}t}\frac{\partial L}{\partial
(\partial\dot\phi_k)}-\partial_j[\frac{\partial L}{\partial
(\partial_j\phi_k)}]\}\delta\phi_k\textrm{d}v(x^k)\textrm{d}t\nonumber\\
&+&\int_{t_0}^{t_1}\int_{\partial\Omega}\{\frac{\partial
L}{\partial(\partial_j\phi_k)}n_j+\frac{\partial
\Gamma_0}{\partial\phi_k}-\frac{\textrm{d}}{\textrm{d}t}\frac{\partial
\Gamma_0}{\partial\dot\phi_k}-\partial_A[\frac{\partial
\Gamma_0}{\partial(\partial_A\phi_k)}]\}\delta\phi_k\textrm{d}a(x^k)\textrm{d}t\nonumber\\
&-&\int_{t_0}^{t_1}\int_{\partial\Omega}2H\{\frac{\partial\hat{\Gamma}}{\partial\phi_k}-\frac{\textrm{d}}{\textrm{d}t}
\frac{\partial\hat{\Gamma}}{\partial\dot\phi_k}-\partial_A[\frac{\partial
\hat{\Gamma}}{\partial(\partial_A\phi_k)}]\}\delta\phi_k\textrm{d}a(x^k)\textrm{d}t+
\int_{t_0}^{t_1}\int_{\partial\Omega}2\frac{\partial
\hat{\Gamma}}{\partial(\partial_A\phi_k)}\partial_AH\delta\phi_k\textrm{d}a(x^k)\textrm{d}t,\end{eqnarray}      
where $n_k$ denotes the unit normal vector on $\partial\Omega$. The
Hamilton's principle asserts that $\delta A[\phi_k]=0$. Therefore,
according to the
fundamental lemma of variation, we have\\
Euler-Lagrange equation:
\begin{equation}\label{s5}\frac {\partial
L}{\partial \phi_k}-\frac{\textrm{d}}{\textrm{d}t}\frac{\partial
L}{\partial (\partial\dot\phi_k)}-\partial_j[\frac{\partial
L}{\partial(\partial_j\phi_k)}]=0,\quad x^k\in\Omega.\end{equation}                         
Natural boundary condition:
\begin{equation}\label{s6}\frac{\partial L}{\partial
(\partial_j\phi_k)}n_j=\frac{\textrm{d}}{\textrm{d}t}\frac{\partial
\Gamma_0}{\partial\dot\phi_k}+\partial_A[\frac{\partial\Gamma_0}{\partial
(\partial_A\phi_k)}]-\frac{\partial\Gamma_0}{\partial
\phi_k}+2H\{\frac{\partial\hat{\Gamma}}{\partial
\phi_k}-\frac{\textrm{d}}{\textrm{d}t}
\frac{\partial\hat{\Gamma}}{\partial\dot\phi_k}-\partial_A[\frac{\partial\hat{\Gamma}}{\partial
(\partial_A\phi_k)}]\}-2\frac{\partial
\hat{\Gamma}}{\partial(\partial_A\phi_k)}\partial_AH,\quad x^k\in\partial\Omega.\end{equation}             
Eq.(\ref{s5}) and (\ref{s6}) show that the surface Lagrangian has no
influence on the Euler-Lagrange equation, but it contributes to the
natural boundary condition and causes the natural boundary condition
to be correlated with the mean curvature and its gradient of
boundary surface. As a boundary condition, Eq.(\ref{s6}) is
universal but complicated. Next, we turn to simplification to
Eq.(\ref{s6}).

\section{Simplification of boundary condition: Tolman's formula}
\label{sec:2}In the classical theory of partial differential
equation, the boundary conditions usually exhibit two features: (1)
they are rate-independent; and (2) they have lower order derivatives
than differential equations themselves. If such two features are
inherited in Eq.(\ref{s6}),
$\Gamma_0(\phi_k,\dot\phi_k,\partial_A\phi_k)$ and
$\hat\Gamma(\phi_k,\dot\phi_k,\partial_A\phi_k)$ necessarily take
the form below
\begin{equation}\label{p1}\Gamma_0(\phi_k,\dot\phi_k,\partial_A\phi_k)=\bar\gamma(\phi_k)+
\bar\chi^A\partial_A\gamma_0(\phi_k)+\chi^{Ak}\partial_A\phi_k.\end{equation}                      
\begin{equation}\label{p2}\hat\Gamma(\phi_k,\dot\phi_k,\partial_A\phi_k)=\hat\gamma(\phi_k)+
\hat\kappa^A\partial_A\gamma_1(\phi_k)+\kappa^{Ak}\partial_A\phi_k.\end{equation}                  
Substituting Eq.(\ref{p1}) and (\ref{p2}) into (\ref{s6}) leads to
\begin{equation}\label{p3}\frac{\partial
L}{\partial(\partial_j\phi_k)}n_j=-\frac{\partial\bar{\gamma}}{\partial
\phi_k}+\partial_A\chi^{Ak}+2H(\frac{\partial\hat\gamma}{\partial
\phi_k}-\partial_A\kappa^{Ak})-2\kappa^{Ak}\partial_AH,\quad x^k\in\partial\Omega,\end{equation}             
where $\bar{\gamma}(\phi_k)$ and $\hat\gamma(\phi_k)$ are two
surface potential energy density functions, while $\chi^{Ak}$ and
$\kappa^{Ak}$ are two surface stresses conjugated to
$\partial_A\phi_k$. The surface stress $\partial_A\phi_k$ and
$\kappa^{Ak}$ are determined by physical property of boundary
surface. We shall discuss them more fully later on.

Let us set a local coordinate system with the base vectors
$(\mathbf{g}_1, \mathbf{g}_2, \mathbf{g}_3)=(\mathbf{g}_A,
\mathbf{n})$, where $\mathbf{g}_A$ ($A=1,2$) is the the covariant
base vectors corresponding to the curvilinear coordinate on the
surface $\partial\Omega$ and $\mathbf{n}$ the unit normal vector. In
such a coordinate system, Eq.(\ref{p3}) can be expanded into
\begin{equation}\label{p4}\frac{\partial
L}{\partial(\partial_j\phi_k)}n_j={\chi^{Ak}}_{,A}+\chi^{Bk}\Gamma_{BA}^A+\chi^{Aj}\Gamma_{Aj}^k-
\frac{\partial\bar\gamma}{\partial\phi_k}+2H(\frac{\partial\hat\gamma}{\partial\phi_k}-
{\kappa^{Ak}}_{,A}-\kappa^{Bk}\Gamma_{BA}^A-\kappa^{Aj}\Gamma_{Aj}^k)-
2\kappa^{Ak}\partial_AH,\quad x^k\in\partial\Omega,\end{equation}                                       
where $\Gamma_{BA}^A$ and $\Gamma_{Aj}^k$ are the connection
coefficients of the surface $\partial\Omega$. The index $k$ takes
$C$ ($C=1,2$) and 3, respectively. Eq.(\ref{p4}) is transformed into
\begin{eqnarray}\label{p5}\frac{\partial
L}{\partial(\partial_j\phi_C)}n_j&=&{\chi^{AC}}_{,A}+\chi^{BC}\Gamma_{BA}^A+\chi^{AB}\Gamma_{AB}^C
-\chi^{A3}b_A^C-\frac{\partial\bar\gamma}{\partial\phi_C}\nonumber\\&+&
2H(\frac{\partial\hat\gamma}{\partial\phi_C}-{\kappa^{AC}}_{,A}-\kappa^{BC}\Gamma_{BA}^A-\kappa^{AB}\Gamma_{AB}^C
+\kappa^{A3}b_A^C)-2\kappa^{AC}\partial_AH,\quad x^k\in\partial\Omega,\end{eqnarray}                                               
\begin{equation}\label{p6}\frac{\partial
L}{\partial(\partial_j\phi_3)}n_j={\chi^{A3}}_{,A}+\chi^{B3}\Gamma_{BA}^A+\chi^{AB}b_{AB}-
\frac{\partial\bar\gamma}{\partial\phi_3}+2H(\frac{\partial\hat\gamma}{\partial\phi_3}-
{\kappa^{A3}}_{,A}+\kappa^{B3}\Gamma_{BA}^A+\kappa^{AB}b_{AB})
-2\kappa^{A3}\partial_AH,\quad x^k\in\partial\Omega,\end{equation}   
where $b_A^C$ is the curvature tensor of the surface
$\partial\Omega$. If the vector field is homogeneous, $\chi^{Ak}$
and $\chi^{Ak}$ are constant tensors. Then, Eq.(\ref{p5}) and
(\ref{p6}) reduce to
\begin{equation}\label{p7}\frac{\partial
L}{\partial(\partial_j\phi_C)}n_j=\chi^{BC}\Gamma_{BA}^A+\chi^{AB}\Gamma_{AB}^C
-\chi^{A3}b_A^C-\frac{\partial\bar\gamma}{\partial\phi_C}+
2H(\frac{\partial\hat\gamma}{\partial\phi_C}-\kappa^{BC}\Gamma_{BA}^A-\kappa^{AB}\Gamma_{AB}^C
+\kappa^{A3}b_A^C)-2\kappa^{AC}\partial_AH,\quad x^k\in\partial\Omega,\end{equation}                       
\begin{equation}\label{p8}\frac{\partial
L}{\partial(\partial_j\phi_3)}n_j=\chi^{B3}\Gamma_{BA}^A+\chi^{AB}b_{AB}-
\frac{\partial\bar\gamma}{\partial\phi_3}+2H(\frac{\partial\hat\gamma}{\partial\phi_3}+
\kappa^{B3}\Gamma_{BA}^A+\kappa^{AB}b_{AB})-2\kappa^{A3}\partial_AH,\quad x^k\in\partial\Omega,\end{equation}      
Furthermore, if the field is also isotropic, we have
$\chi^{AB}=\sigma g^{AB}$, $\chi^{A3}=0$, $\kappa^{AB}=\tau g^{AB}$
and $\kappa^{A3}=0$. Therefore, Eq.(\ref{p7}) and (\ref{p8}) lead to
\begin{equation}\label{p9}\frac{\partial L}{\partial(\partial_j\phi_C)}n_j=
\sigma{g^{BC}}_{,B}-\frac{\partial\bar\gamma}{\partial\phi_C}+
2H(\frac{\partial\hat\gamma}{\partial\phi_C}-\tau{g^{BC}}_{,B})-2\tau\partial^CH,\quad x^k\in\partial\Omega.\end{equation}     
\begin{equation}\label{p10}\frac{\partial L}{\partial(\partial_j\phi_3)}n_j=2\sigma
H-\frac{\partial\bar\gamma}{\partial\phi_3}+2H(\frac{\partial\hat\gamma}{\partial\phi_3}-2\tau
H),
\quad x^k\in\partial\Omega.\end{equation}                                                                     
Here, $\sigma$ and $\tau$ are two surface tensions, and $g^{AB}$ is
the metric tensor of surface. It is easy to see that Eq.(\ref{p10})
can be equivalently represented as
\begin{equation}\label{p11}\frac{\partial L}{\partial(\partial_j\phi_k)}n_jn_k=2\sigma
H-\frac{\partial\bar\gamma}{\partial\phi_k}n_k+2H(\frac{\partial\hat\gamma}{\partial\phi_k}n_k-2\tau
H),\quad x^k\in\partial\Omega.\end{equation}                                                                  

Consider a liquid droplet. Let $\phi_k$ is a displacement field.
Because the surface potential energies are invariant under the
transformation of rigid motion, it is necessary that
$\bar{\gamma}(\phi_k)$ and $\hat\gamma(\phi_k)$ are independent of
$\phi_k$. As a result, Eq.(\ref{p9}) and (\ref{p11}) reduce to
\begin{equation}\label{p12}\frac{\partial L}{\partial(\partial_j\phi_C)}n_j=
\sigma{g^{BC}}_{,B}-2\tau H{g^{BC}}_{,B}-2\tau\partial^CH,\quad x^k\in\partial\Omega.\end{equation}           
\begin{equation}\label{p13}\frac{\partial L}{\partial(\partial_j\phi_k)}n_jn_k=2\sigma
H-4\tau H^2,\quad x^k\in\partial\Omega.\end{equation}                                                         
In physics, the right-side term of Eq.(\ref{p13}) represents the
pressure, denoted by $\Delta p$. Let $\delta=2\tau/\sigma$. Clearly,
it has the dimension of length. As thus, Eq.(\ref{p13}) is rewritten
as
\begin{equation}\label{p14}\Delta p=2\sigma
H(1-\delta H),\quad x^k\in\partial\Omega.\end{equation}                                                         
Eq.(\ref{p14}) is just the Tolman's formula \cite{11}. It has been
extensively applied to analyze the surface size effects of
micro/nano-scale liquid droplet and solid particle \cite{12,13}.

Interestingly, if we assume $\hat\Gamma=\delta\Gamma_0/2$,
Eq.(\ref{s6}) will lead to
\begin{equation}\label{p15}\frac{\partial L}{\partial
(\partial_j\phi_k)}n_j=(1-\delta
H)(\frac{\textrm{d}}{\textrm{d}t}\frac{\partial
\Gamma_0}{\partial\dot\phi_k}+\partial_A[\frac{\partial\Gamma_0}{\partial
(\partial_A\phi_k)}]-\frac{\partial\Gamma_0}{\partial\phi_k})-\frac{\partial
\Gamma_0}{\partial(\partial_A\phi_k)}\partial_A(\delta H),\quad x^k\in\partial\Omega.\end{equation}         
Eq.(\ref{p15}) can be regarded as a extension of the Tolman formula.

\section{Conclusion}
\label{sec:2}In the framework of the Lagrangian field theory, we
propose the so-called surface Lagrangian to characterize the surface
effects of field, The surface Lagrangian has no influence on the
Euler-Lagrange equation, but it contributes to the natural boundary
condition and causes a type of new non-classical natural boundary
condition to be correlated with the mean curvature of boundary
surface. The well-known Tolman's formula is derived from
simplification to this new natural boundary condition.

\section*{Acknowledgements}
The support of the National Nature Science Foundation of China
through the Grant No. 11172130 is gratefully acknowledged.

\end{document}